\documentclass[prb,
twocolumn,
superscriptaddress,showpacs,amsmath,amssymb]{revtex4}

\usepackage{xcolor}
\usepackage{graphicx}

\begin{document}

\title{Gravity from symmetry breaking phase transition}

\author{G.E.~Volovik}
\affiliation{Low Temperature Laboratory, Aalto University,  P.O. Box 15100, FI-00076 Aalto, Finland}
\affiliation{Landau Institute for Theoretical Physics, acad. Semyonov av., 1a, 142432,
Chernogolovka, Russia}

In memory of Dmitry Diakonov

\date{\today}

\begin{abstract}
{The paper  is devoted to the memory of Dmitry Diakonov. We discuss gravity emerging in the fermionic vacuum as suggested by Diakonov 10 years ago in his paper "Towards lattice-regularized Quantum Gravity".\cite{Diakonov2011}
 Gravity emerges in the phase transition. The order parameter in this  transition is the tetrad field, which appears as the bilinear composite of the fermionic fields. This symmetry breaking gives 6 Nambu-Goldstone modes;  6 gauge bosons in the spin-connection fields, which absorb 6 NG modes and become massive gauge bosons; and  6 Higgs fields. The similar scenario of the symmetry breaking with appearance of the $5+1$ Higgs gravitons takes place in the B-phase of superfluid $^3$He. While in $^3$He-B these Higgs modes are all massive, in the GR these Higgs collective modes  give rise to two massless gravitational waves.
}
\end{abstract}
\pacs{
}

\maketitle

\section{Introduction}

Let us follow the reasoning of  Dmitry Diakonov,\cite{Diakonov2011} that the gravitational tetrad fields emerge as bilinear combinations of the fermionic fields in the symmetry breaking phase transition. The picture of such phase transition should contain the order parameter, "superfluid" stiffness, Ginzburg-Landau functional, Bogoliubov-de Gennes fermionic quasiparticles, Nambu-Goldstone modes, topological defects, Higgs modes, massive gauge bosons, etc. We discuss these issues using the analogy with symmetry breaking scheme in superfluid $^3$He-B. The analogy suggests that the gravitational waves in general relativity emerge as the Higgs amplitude modes, i.e. the gravitons are Higgs bosons.  But as distinct from $^3$He-B these gravitons are massless.

\section{Symmetry breaking}
\label{Breaking}

The extension of the symmetry breaking scheme, which takes place in superfluid $^3$He-B,\cite{Leggett1973,VollhardtWolfle,Volovik2003} to the relativistic vacuum suggests the following symmetry breaking scenario.\cite{Volovik1990}
  The original deep quantum vacuum  is assumed  to be symmetric with respect to the high group $G$, such as in  Refs.\cite{MacDowell1977,Townsendl1977,Verwimp1990}.  According to the scenario suggested by $^3$He-B, the symmetry group of the original relativistic vacuum  $G$ is the product of two independent Lorentz groups: the vacuum has the symmetry under the  Lorentz space-time rotations $L_{\rm c}$ (which can be extended to the group of coordinate transformations), and the symmetry under the Lorentz  3+1 transformations of the fermionic spins $L_{\rm s}$. 
  
  The original symmetric vacuum does not contain gravity. The fundamental variables in this vacuum are the fermion fields  $\Psi$ and the gauge field $\omega_{\mu}^{bc}$ --  the connections of the Lorentz group $L_{\rm s}$.\cite{Diakonov2011} Gravity appears after
this vacuum experiences the symmetry breaking of these two Lorentz symmetries to their diagonal subgroup: 
\begin{equation}
G= L_{\rm c} \times L_{\rm s} \rightarrow H=L_{J} \,.
\label{BrokenSymmetry4}
\end{equation}
The order parameter of this symmetry breaking is the matrix $e^\mu_a$, which plays the role of the tetrad  of the emergent $3+1$ gravity. 
The Minkowski vacuum becomes degenerate with respect to each of the two $SO(1,3)$ rotations, but remains invariant under the combined Lorentz transformation  $L_{J}$. The latter represents the rest symmetry $H$ of the present degenerate vacuum.

In the original symmetric vacuum, the spin rotation group $L_{\rm s}$ is local, while the space rotation group $L_{\rm c}$ is global. 
Since in the symmetric vacuum the metric is absent, the distance between the space-time points is not determined. This shows that this vacuum is chaotic, but is isotropic.  The $L_{\rm c}$ symmetry of the original action gives rise 
to the symmetry under coordinate transformations with fixed determinant.

\section{Model for symmetric vacuum}
\label{SymmetricVac}

We consider the simplest model illustrating this symmetry breaking pattern $G\rightarrow H$. Following Refs.\cite{Diakonov2011,Wetterich2003,Diakonov2012,Vergeles2021,Nesti2010,Percacci2018}, let us choose the following action on the fundamental microscopic scale,  which depends only on the fundamental variables   $\Psi$ and  $\omega_{\mu}^{bc}$:
 \begin{equation}
S_{\rm micro}= \frac{1}{24}  e_{abcd}\int  \Theta^a\wedge \Theta^b \wedge \Theta^c\wedge \Theta^d \,,
\label{micro}
\end{equation}
where
\begin{equation}
 \Theta^a = \frac{i}{2} \left[ \bar\Psi \gamma^a D_\mu \Psi - D_\mu \bar\Psi\gamma^a \Psi\right] dx^\mu
\,.
\label{Theta}
\end{equation}
For simplicity, we consider a single Dirac fermion. But the extension to the more complicated vacua is straightforward.
Eq.(\ref{micro}) contains 8 fermion fields, which is very different from the typical theories, that are quadratic in fermionic fields.
The conventional quadratic action for the Weyl and/or Dirac fermions appears only after the symmetry breaking, when the tetrads and metric emerge.

We assume that there is no $(\Gamma_\mu)^\rho_\sigma$-fields, which describe the geometrical Levi-Civita gravitational connection, i.e. there are no dynamical fields related to the space-time rotations and coordinate transformations. In other words, the group of space-time rotations is global. This is natural in the theory, where gravity is the emergent phenomenon. Invariance  of Eq.(\ref{micro})  under the Lorentz group of the space-time rotations automatically induces the  invariance  of Eq.(\ref{micro}) under coordinate transformations. Such induced  invariance will be automatically extended to the actions obtained after symmetry breaking, though maybe restricted by the coordinate transformations with conserved determinant -- the unimodular general coordinate transformations.

In  Eq.(\ref{micro}), the fermionic field $\Psi$  has mass dimension zero, which is consistent with the lattice model of the quantum vacuum.\cite{Volovik2021b} The symmetry with respect to the coordinate transformations is absent in the lattice theories, and may appear only after coarse graining at low energies. The more complicated models should take into account the chirality of fermions.\cite{Zubkov2021}

The action  (\ref{micro}) has the full symmetry $G$ of the microscopic vacuum without gravity. Since the metric is absent, the distance between the spacetime points is not determined, as a result the original vacuum is not sensitive to the coordinate transformations. The spacetime metric and the rigidity of the vacuum appear only after the symmetry breaking. This also means that the action for the Standard Model gauge fields $F_{\mu\nu}$ is also absent, except for the topological terms, which do not depend on metric field. The vector gauge fields $A_\mu$ enter only the fermionic action via the corresponding long derivative $D_\mu$. The action for the field strength
 $F_{\mu\nu}$, which appears after symmetry breaking, is obtained by integration over fermions.

The original vacuum state with action Eq.(\ref{micro}) has the nontrivial topology, which supports the formation of massless fermions, see Sec. \ref{Tetrad}.
The original action, which contains only the differential forms, may also include the fundamental 4-form field $F_{\mu\nu\alpha\beta}$, which was introduced by Hawking to describe the vacuum energy,\cite{Hawking1984} and used for the solution of the cosmological constant problem.\cite{KlinkhamerVolovik2008a,KlinkhamerVolovik2008,KlinkhamerVolovik2021} This field is the totally antisymmetric pseudo-tensor, which does not depend on metric. The 4-form field may also emerge as the order parameter together with the tetrad fields.\cite{Nesti2010,Percacci2018}

\section{Spin connection action}
\label{HigherOrder}

The original symmetric vacuum  has also the spin connection degrees of freedom, which interact with fermions in Eq.(\ref{micro}) via the long derivative $D_\mu$:
\begin{equation}
D_\mu = \partial_\mu + \frac{1}{8} \omega_{\mu bc} \left[ \gamma^b, \gamma^c \right]\,.
\label{LongDerivative}
\end{equation}
Here we do not discuss the vector gauge fields which also enter the long derivative $D_\mu$.
 The corresponding Yang-Mills action for spin connection has the form:
\begin{equation}
S_\omega= K \int R^{ab}\wedge R^{cd} \epsilon_{abcd} \,,
\label{SpinConnectionTerm}
\end{equation}
where the curvature -- the  strength of the  Yang-Mills field --  is
\begin{equation}
R^{ab}_{\mu\nu}  = \partial_\mu \omega^{ab}_\nu - \partial_\nu \omega^{ab}_\mu
+ \omega^{a}_{c\mu}\omega^{cb}_\nu -  \omega^{a}_{c\nu}\omega^{cb}_\mu\,.
\label{Curvature}
\end{equation}
The curvature contains the tensor $\eta_{ab}$.

The parameter $K$ is dimensionless. It is not excluded that it is quantized, because the Eq.(\ref{SpinConnectionTerm}) is the full derivative and belongs to Pontryagin class. However, since the fermionic field is dimensionless, the Eq.(\ref{SpinConnectionTerm}) can be extended by including the scalar $\bar\Psi \Psi$, which makes the term dynamical
\begin{equation}
S \propto  \int \bar\Psi\Psi \, R^{ab}\wedge R^{cd} \epsilon_{abcd} \,.
\label{SpinConnectionTermPsi}
\end{equation}
The propagating modes of this gauge field will be discussed in Sec.\ref{Higgs}.
These gauge bosons obtain masses after symmetry breaking.

\section{Tetrad as the order parameter of the symmetry breaking phase transition}
\label{Tetrad}

In the Diakonov scenario, the quantities $\Theta^a$ in the symmetric phase play the role of the virtual tetrads. The fermionic action (\ref{micro}) can be rewritten as\cite{Diakonov2011}
 \begin{equation}
S_{\rm M}= \frac{1}{6}e_{abcd}\int \Theta^a\wedge \hat e^b \wedge \hat e^c\wedge \hat e^d \,,
\label{VirtualTetrad}
\end{equation}
where
 \begin{equation}
\hat e^a = \Theta^a \,.
\label{VirtualTetrad2}
\end{equation}
In this form, the action  describes the massless Weyl or Dirac fermions in the background of the virtual tetrads $\hat e^a$. The gapless spectrum of fermions is protected by the $\pi_3$ topological invariant,\cite{Volovik2003} which characterizes Weyl and Dirac nodes in the fermionic spectrum. In principle, the topology can be the starting point of the  action (\ref{micro}).

The tetrad fields become physical after the symmetry breaking. The symmetry breaking is represented by the appearance of the nonzero Higgs field -- the Bogoliubov quasi-average -- the vacuum expectation value (vev) of the bilinear fermionic field:\cite{Akama1978,Diakonov2011,Diakonov2012,ObukhovHehl2012,Wetterich2021,Volovik2021c}
 \begin{equation}
e^a =\left< \Theta^a\right> \,.
\label{quasiaverage}
\end{equation}
This order parameter $e^a_\mu$ is the emerging physical tetrad field, which fluctuates around the nonzero vev, rather than around 0 in symmetric vacuum. 

The tetrad $e^a_\mu$ transforms as a derivative and thus has mass dimension 1, which leads to the dimensionless interval. This is distinct from the GR, where  the metric has zero dimension and the interval has dimension of length. On the other hand the fermionic field $\Psi$ has zero dimension, see discussion in Ref.\cite{Diakonov2012}. This leads to physics without dimensions, i.e. where all the diffeomorphism invariant quantities are dimensionless, see also review \cite{Volovik2021b}.

The action for the order parameter tetrad field -- the analog of the Ginzburg-Landau functional -- contains two main terms: zeroth order and second order in gradients.

\section{Vacuum energy and cosmological constant}
\label{Lambda}

The zeroth order term in the expansion gives the gravitational term corresponding to the cosmological constant:
 \begin{equation}
S_0= \frac{\Lambda}{24}  e_{abcd}\int e^a\wedge e^b \wedge e^c \wedge e^d \,.
\label{ZerothOrder}
\end{equation}
The cosmological constant $\Lambda$ has zero dimension, and thus in principle can be quantized in integer numbers. In such scenario, the zero value of $\Lambda$ is natural.

In the other scenario, instead of the constant $\Lambda$, one uses the function of the tetrad determinant $\epsilon(e)$:
 \begin{equation}
S_0\{e\}= \frac{1}{24}  e_{abcd}\int  \, \epsilon(e) \,e^a\wedge e^b \wedge e^c \wedge e^d \,.
\label{ZerothOrderLambda}
\end{equation}
Then the cosmological "constant", i.e. the vacuum energy density, which enters Einstein equations,  will be:\cite{Volovik2021c}
 \begin{equation}
\Lambda(e)\equiv \rho_{\rm vac}(e) =\epsilon(e) + e \frac{d\epsilon}{de}\,.
\label{rho_vac}
\end{equation}
The equation (\ref{ZerothOrderLambda}) gives rise to the equilibrium value $e_{\rm eq}$ of the tetrad determinant, which satisfies the equation
 $\Lambda(e_{\rm eq})=0$. This automatically leads to the nullification of the cosmological constant in the fully equilibrium vacuum without matter. This scenario of the nullification of the cosmological constant in equilibrium is similar to that in Refs.\cite{KlinkhamerVolovik2008a,KlinkhamerVolovik2008}, where the 4-form field $F_{\mu\nu\alpha\beta}$ is used instead of the determinant $e$. The other vacuum variable in terms of two types of tetrads, geometric and crystalline, is in Ref.\cite{KlinkhamerVolovik2019}.
 
The function $\epsilon(e)$ explicitly violates symmetry under coordinate transformations, but the invariance under coordinate transformations with fixed determinant is preserved -- the unimodular general coordinate transformations. The further extension would be to use two determinants, say, for left and right fermions,\cite{Volovik2021c} see also Ref.\cite{ParhizkarGalitski2021}.

\section{Vacuum stiffness}
\label{Stiffness}

The gradient term in the order parameter action is the standard Palatini action of general relativity:
\begin{equation}
S_{\rm grad}= \frac{1}{G}\int e^a\wedge e^b\wedge R^{cd} \epsilon_{abcd} \,.
\label{action_grad}
\end{equation}
Connections $\omega^a_b$ and tetrads $e^a$ have dimension 1, the curvature $R^{cd} $ has dimension 2, and thus the parameter $1/G$ has zero dimension.

Variation over connections $\omega^a_b$ in the absence of matter and ignoring the term (\ref{SpinConnectionTerm}) gives:
\begin{equation}
D(e^{[a}\wedge e^{b]})=0 \,.
\label{NoTorsion}
\end{equation}
So,  in the absence of matter and for $K=0$ in Eq.(\ref{SpinConnectionTerm}) there is no torsion.
Eq.(\ref{NoTorsion})  can be solved for the 24 connection fields in terms of 16 tetrads (the other components of connection are related to torsion). As a result this term contains the quadratic form of the gradients of tetrads.
That is why the prefactor $1/G$ plays the role of the superfluid stiffness in superfluids and superconductors, and can be called the gravitational stiffness, which determines the elasticity of spacetime -- "the metrical elasticity".\cite{Sakharov1967} 

In principle, $1/G$ can be also the function of $e$:
\begin{equation}
S_{\rm grad}= \int \frac{1}{G(e)} e^a\wedge e^b\wedge R^{cd} \epsilon_{abcd}.
\label{action_grad_e}
\end{equation}
Following Refs. \cite{KlinkhamerVolovik2008a,KlinkhamerVolovik2008} for the phenomenology of the vacuum variable, one may suggests that $1/G$ may have the form:
\begin{equation}
\frac{1}{G}=\frac{e^2}{e_{\rm eq}^2} \,.
\label{G_e}
\end{equation}
In this approach, the gravitational stiffness is absent in the symmetric vacuum, where $e=0$. However, this is not necessary, since the action (\ref{action_grad}) anyway disappears when $e^a=0$. That is why the gravitational stiffness $1/G$ can be kept constant. Moreover, it is possible that $1/G$ is the topological invariant in the 4D crystals, similar to the other topological invariants in the Brillouin zone.\cite{NissinenVolovik2018b,NissinenHeikkilaVolovik2021,Nissinen2020} And again, $G(e)$ is symmetric only under the unimodular general coordinate transformations.

\section{NG modes, tetrad Higgs and gravitons}
\label{Higgs}

In $^3$He-B there are 18 components of the order parameter, 9 real and 9 imagrinary.\cite{VollhardtWolfle} They can be rearranged in terms of two tetrad fields: for left and for right fermions.\cite{Volovik1990}  Here we consider only $3\times 3=9$ real components of the order parameter, which corresponds to the same tetrad for left and right fermions. The corresponding symmetry breaking has the form $SO(3)_s\times SO(3)_c \rightarrow SO(3)_J$. In this symmetry breaking there are $9=3+6$ collective modes: 3 Nambu-Goldstone (NG) modes and 6 Higgs amplitude modes.\cite{VolovikZubkov2014,Sauls2017}
If the $SO(3)_s$ spin rotation group is gauged, all the  NG modes become massive gauge bosons.

The 6 collective Higgs amplitude modes in $^3$He-B represent $1+5$ massive gravitational waves.\cite{Volovik1990,Volovik1992} 
The Higgs mode with $J=0$ describes the oscillation of $\delta e^i_i$. This is the so-called pair-breaking mode.\cite{VollhardtWolfle} The other five Higgs modes with $J=2$ are $\delta e^i_k + \delta e^k_i - \frac{2}{3} \delta^i_k \delta e^l_l$. These are the so-called real squashing modes.\cite{VollhardtWolfle} Masses of these 6 "gravitons" have been measured, see e.g. recent experiments.\cite{Halperin2019}

These massive modes are similar to gravitons in massive GR.\cite{Zakharov1970} This similarity is not accidental, since $^3$He-B can be extended to the relativistic medium.\cite{SilaevVolovik2010} However, when extending  $^3$He-B to the relativistic system,  one should take into account that the breaking of symmetry, which contains time and gauge, requires different approach to the counting of the NG modes,\cite{NielsenChadha1976} see  the modes in ferromagnets,\cite{VolovikZubkov2014} the review \cite{Watanabe2020} and the recent paper \cite{Hidaka2021}. 

In the relativistic vacuum of the Diakonov model, we have 6 spin connection gauge fields $\omega^{ab}_\mu$ of the local symmetry group $L_s$. The counting scheme of the dynamic modes of spin connection follows the same procedure as in the case of the vector gauge field $A_\mu$. The $A_0$ component should be ignored, since it is related  to the space-time Lorentz boost, and one more constraint follows from the gauge invariance leaving only two modes of the propagating photons. In the same manner, in case of spin connection gauge fields, 6 components  $\omega^{ab}_0$ are excluded, the other 6 DoF are excluded due to gauge invariance. But when the symmetry is spontaneously broken according to Eq.(\ref{BrokenSymmetry4}), all the 6 gauge bosons absorb 6 NG modes and become massive. 

In the same manner the time components $e^a_0$ should be excluded from the $4\times 4=16$  components of the tetrad order parameter, since they are related  to the space-time Lorentz boost, which enters the group $L_c$.   The rest $12=6+6$ degrees of freedom  $e^a_i$ include 6 NG modes, which are absorbed by gauge bosons. 

The other 6 tetrad modes are  the Higgs modes. As in $^3$He-B, the Higgs modes represent $1+5$ gravitons with $J=0$ and $J=2$ correspondingly. This is different from theories, where gravitons are considered as NG  bosons.\cite{Phillips1966,Chkareuli2011}  

In $^3$He-B, all $1+5$ gravitons are massive. In relativistic case the situation is different, since the symmetry under coordinate transformations becomes important. In case of symmetry under unimodular general coordinate transformations, only  the $J=0$ graviton remains massive, as in $^3$He-B. From the  5 components of the  Higgs gravitons with $J=2$, three degrees of freedom can be removed by the transformation of the spatial coordinates.   The other two degrees of freedom with $J_z=\pm 2$ represent massless gravitational waves,  as in GR. 

In the above approach we took into account that there is no independent $(\Gamma_\mu)^\rho_\sigma$-fields, which describe the gravitational connection as the gauge field, i.e. there are no dynamical gauge fields related to $L_c$ and to the coordinate transformations. In other words, the group of coordinate transformations is "global". This is natural in the theory, where gravity is the emergent phenomenon.

Massive gravity in terms of tetrads has been discussed in Ref.\cite{Hinterbichler2012}. It will be interesting to discuss the  van Dam-Veltman-Zakharov (vDVZ) discontinuity\cite{Zakharov1970,vanDam1970} in this approach.  It will be also interesting to check whether the massive graviton with zero helicity, $J=0$, experiences the Boulware-Deser ghost instability in the Diakonov model.

\section{Matter as quasiparticles in the gravitational background}
\label{Matter}

As follows from Eq.(\ref{micro}), the action for quasiparticles in the background of the order parameter $e^a$  (the matter action) has the form:
\begin{equation}
S_{\rm M}= \frac{1}{6}e_{abcd}\int \Theta^a\wedge e^b \wedge e^c\wedge e^d
\,.
\label{ActionM}
\end{equation}
This shows that fermionic degrees of freedom have the form of the Dirac or Weyl fermionic quasiparticles interacting with effective gravity described by the Einstein-Cartan-Sciama-Kibble type theory.
Eq.(\ref{ActionM}) is the analogue of the Bogoliubov--de Gennes action for quasiparticles in superfluids and superconductors.

The massless Weyl and Dirac fermions represent the topological matter. The topological invariants for the matter field emerge after symmetry breaking.  However, this topology was already hidden in the original symmetric vacuum in Eq.(\ref{VirtualTetrad}).

 Let us mention that here we considered the $^3$He-B scenario of  gravity emerging together with  massless Weyl and Dirac fermions. 
This scenario is based on the symmetry breaking of two Lorentz groups to their diagonal subgroup. Another scenario of emergent gravity and  massless Weyl fermions is suggested by another phase of superfluid $^3$He, the chiral superfluid $^3$He-A. The $^3$He-A scenario is purely topological: gravity and  Weyl fermions emerge together with the Lorentz invariance in the vicinity of the topologically protected crossing points in the fermionic spectrum -- the Weyl points.\cite{Volovik2003,Horava2005}

 \section{Degenerate Minkowski vacua and topological objects}
 \label{defects}

The homotopy groups of the manifold of the degenerate Minkowski vacua are
\begin{eqnarray}
 \pi_n(G/H)=\pi_n(O(3,1))
 =\pi_n(SO(3,1)\times Z_2\times Z_2) \,.
\label{nL}
\end{eqnarray}
Here we took into account that formation of the tetrad field also breaks the discrete symmetries.\cite{Vergeles2021}
The parity symmetries, $P_{\rm c}$ and $P_{\rm s}$, are broken to the diagonal subgroup: $P_{\rm c}  \times  P_{\rm s}\rightarrow P$, where  $P$ is the combined discrete symmetry. The same is with the time reversal: $T_{\rm c}  \times  T_{\rm s} \rightarrow T$, where $T$ is the combined symmetry operation. 

The group $\pi_0(G/H)=Z_2\times Z_2$ gives rise to three types of cosmic domain walls, in which either the space components of tetrads  change sign, or time component changes sign, or the whole tetrad changes sign.   In $^3$He-B these walls are non-topological due the extra degrees of freedom of the order parameter.\cite{SalomaaVolovik1988}  
They are the analogs of the Kibble-Lazarides-Shafi cosmic walls bounded by strings.\cite{Kibble1982,Makinen2019} 

The fundamental group of the manifold of the Minkowski vacuum states is $\pi_1(G/H)=Z_2$.  The nontrivial element of the group describes the torsion string, in which the curvature has $\delta$-function singularity on the axis.\cite{Volovik2021c}
  In $^3$He-B it is the $Z_2$ spin vortex.\cite{Kondo1992}  

 The other topological objects in tetrad gravity are the topological instantons.\cite{HansonRegge,AuriaRegge}
In principle, the Big Bang can be also considered as the topological defect, in which the tetrads\cite{Sakharov1980,Turok2018,Turok2021}   or topological invariants\cite{KlinkhamerVolovik2021}  change sign.

\section{Conclusion}
\label{ConclusionSec}

Here, following the approach of Dmitri Diakonov, we discussed the symmetry breaking phase transition, which leads to the formation of gravity. Similar to the phase transition in superfluid $^3$He-B, the role of the order parameter is played by the tetrad fields. This symmetry breaking gives: 

(i)  6 Nambu-Goldstone modes; 

(ii)  6 gauge bosons in the spin-connection fields, which absorb 6 NG modes and become massive gauge bosons;

(iii) 6 Higgs collective modes, which are  $5+1$ gravitons.  In massive GR these massive gravitons are similar to the Higgs modes in $^3$He-B.  In the GR with symmetry under unimodular coordinate transformations,  they give massive $J=0$ graviton and  two conventional gapless gravitational waves with $J_z=\pm 2$.

 In conclusion, there are many interesting consequences of the Diakonov theory of emergent gravity, including
the gravitational waves emerging as the Higgs bosons of the symmetry breaking phase transition; physics without dimensions where all the diffeomorphism invariant quantities are dimensionless; absence of the interval in the original symmetric vacuum; solution of the cosmological constant problem; etc. All this and also the analogy with superfluid $^3$He-B indicate the advisability of the further development of the Dyakonov model for the construction of a consistent theory of quantum gravity.

 {\bf Acknowledgements}. 
I thank F. Klinkhamer, S. Vergeles and M. Zubkov for discussions.
  This work has been supported by the European Research Council (ERC) under the European Union's Horizon 2020 research and innovation programme (Grant Agreement No. 694248).

\end{document}